\documentclass[9pt]{IEEEtran}

\usepackage{amsfonts}
\usepackage{amsmath}
\usepackage{amssymb}
\usepackage{mathrsfs}
\usepackage{amstext}

\font\msbm=msbm10

\newtheorem{Theorem}{Theorem}[section]
\newtheorem{lemma}[Theorem]{Lemma}
\newtheorem{corollary}[Theorem]{Corollary}

\newtheorem{proposition}[Theorem]{Proposition}

\def\mathbb#1{\hbox{\msbm{#1}}}
\newcommand{\N}{{\mathbb{N}}}
\newcommand{\R}{{\mathbb{R}}}

\newcommand{\C}{{\mathbb{C}}}

\renewcommand{\P}{{\mathbb{P}}}
\newcommand{\E}{{\mathbb{E}}}
\newcommand{\A}{{A}}

\newcommand{\sgn}{\operatorname{sgn}}
\newcommand{\Tr}{\operatorname{Tr}}
\newcommand{\sparsity}{s}
\newcommand{\atom}{{a}}

\newcommand{\beq}{\begin{eqnarray}}

\newcommand{\eeq}{\end{eqnarray}}
\newcommand{\beqn}{\begin{eqnarray*}}
\newcommand{\eeqn}{\end{eqnarray*}}

\newcommand{\supp}{\operatorname{supp}}

\newcommand{\qed}{\IEEEQEDclosed}
\newenvironment{Proof}{\noindent
{\bf\underline{Proof:} }}
{\hspace*{\fill}\qed\vskip1em}
\begin{document}
\title{Circulant and Toeplitz Matrices in Compressed Sensing}

\author{\IEEEauthorblockN{Holger Rauhut\\}
\IEEEauthorblockA{Hausdorff Center for Mathematics and Institute for Numerical Simulation\\
University of Bonn,
Endenicher Allee 60, D-53115 Bonn\\
rauhut@hcm.uni-bonn.de\\
}}
\date{\today}


\maketitle

\begin{abstract}
Compressed sensing seeks to recover a sparse vector from a small number of
linear and non-adaptive measurements. While most work so far focuses on
Gaussian or Bernoulli random measurements we investigate the use of partial random
circulant and Toeplitz matrices in connection with recovery by $\ell_1$-minization.
In contrast to recent work in this direction we allow the use of an arbitrary subset
of rows of a circulant and Toeplitz matrix. Our recovery result predicts that the necessary number of measurements to ensure sparse reconstruction by
$\ell_1$-minimization with random partial circulant or Toeplitz matrices scales linearly in the sparsity up to a $\log$-factor in the ambient dimension. 
This represents a significant improvement over previous recovery results for
such matrices. As a main tool for the proofs we use a new version of
the non-commutative Khintchine inequality.
\end{abstract}
%






\section{Introduction}

Compressed sensing is a recent concept in signal processing where one seeks
to reconstruct efficiently a sparse signal from a minimal number of linear and non-adaptive measurements \cite{do06-2}. So far various measurement matrices have been investigated, 
most of them random matrices. Among these are
Bernoulli and Gaussian matrices \cite{badadewa06} (with independent $\pm 1$ or standard normal entries) 
as well as partial Fourier matrices \cite{cata06,ra08-1,ru06-1}. Recently, Bajwa et al.\ \cite{bahanorawr07} 
(see also \cite{ro08}) studied 
Toeplitz type and circulant matrices 
in the context of compressed sensing where the entries of the vector generating the Toeplitz or
circulant matrix are chosen at random according to a suitable
probability distribution. Compared to Bernoulli or Gaussian matrices random 
Toepliz and circulant matrices have the advantage
that they require a reduced number of random numbers to be generated. More importantly, there are
fast matrix-vector multiplication routines which can be exploited in recovery algorithms. Furthermore, they arise naturally in certain applications
such as identifying a linear time-invariant system \cite{bahanora08}.

Basis Pursuit ($\ell_1$-minimization) is one of the major approaches to efficiently recover 
a sparse vector. This technique is quite well understood by now.
Modern optimization algorithms \cite{bova04} such as LARS \cite{efhajoti04} (sometimes called homotopy method) are reasonably fast.

Bajwa et al.\ \cite{bahanorawr07,bahanora08} estimated the so-called restricted isometry constants 
of a random Toeplitz type or circulant matrix which then allows to provide
recovery guarantees for $\ell_1$-minimization. However, their bound is very pessimistic compared
to related estimates for Bernoulli / Gaussian or partial Fourier matrices. More precisely, the estimated
number of measurements grows with the sparsity squared, 
while one would rather expect a linear scaling. 
Indeed, this is also suggested by numerical experiments. 
We close the theoretical gap by providing recovery guarantees 
for $\ell_1$-minimization in connection with circulant and Toeplitz type matrices where the necessary number of
measurements scales linearly with the sparsity. However, we do not make use of 
the restricted isometry constants and a good estimate of the latter is therefore still open.

\section{Sparse recovery with circulant and Toeplitz matrices}

For a vector $x\in \R^N$ we let $\supp x = \{j, x_j \neq 0 \}$ denote its support and
$\|x\|_0 = |\supp x|$ the number of non-zero entries. It is called $\sparsity$-sparse if $\|x\|_0 \leq \sparsity$. 
We aim at recovering $x$ from $y= Ax \in \R^n$ where $A$ is a suitable $n \times N$ measurement
matrix and $n < N$. A natural strategy is to consider $\ell_0$-minimization,
\begin{equation}\label{l0}
\min_x \|x\|_0 \quad \mbox{ subject to } Ax = y. 
\end{equation}
Unfortunately this combinatorial optimization problem is NP hard in general \cite{avdama97}. 
Therefore, we solve instead the convex problem
\begin{equation}\label{l1}
\min \|x\|_1 \quad \mbox{ subject to } Ax=y,
\end{equation}
where the $\ell_p$-norm is defined as usual, $\|x\|_p= (\sum_{j=1}^N |x_j|^p)^{1/p}$.
It is by now well understood 
that the solutions of both minimization problems often coincide
and are equal to the original vector $x$, see e.g.\ 
\cite{carota06,dota06,do06-2,grni03,ra07}.
A by now popular result \cite{carota06,ca08,folaXX} states that indeed 
\eqref{l1} (stably) recovers all $\sparsity$-sparse $x$
from $y = Ax$ provided the restricted isometry constant 
$\delta_{2\sparsity} \leq \delta < \sqrt{2}-1$. The latter means that
\[
(1-\delta) \|x\|_2^2 \leq \|Ax\|_2^2 \leq (1+\delta) \|x\|_2^2
\]
for all $2\sparsity$-sparse vectors $x$. It is known \cite{badadewa06}
that random Gaussian or Bernoulli matrices, i.e. $n\times N$ matrices with independent and normal distributed or Bernoulli distributed entries, satisfy this condition with probability at least $1-\epsilon$ 
provided $\sparsity \leq C_1 n \log(N/\sparsity)  + C_2 \log(\epsilon^{-1})$. 



We consider the following types of measurement matrices. For
$b = (b_0,b_1, \hdots,b_{N-1}) \in \R^N$ we let its associated circulant 
matrix $S = S^b \in \R^{N\times N}$ with entries
$S_{i,j} = b_{j-i \mod N}$, where $i,j =1,\hdots,N$.
Similarly, for a vector $c = (c_{-N+1},c_{-N+2},\hdots,c_{N-1})$ its associated Toeplitz matrix
$T = T ^c \in \R^{N \times N}$ has entries
$T_{i,j} = c_{j-i}$, where $i,j =1,\hdots,N$.
Now we choose an arbitrary subset $\Omega \subset \{1,\hdots,N\}$ of cardinality $n<N$ and let the partial circulant matrix
$S_\Omega = S_\Omega^b \in \R^{n \times N}$ be the submatrix of $S$ consisting of 
the rows indexed by $\Omega$. The partial Toeplitz
matrix $T_\Omega=T_\Omega^c \in \R^{n \times N}$ is defined similarly. In this
paper the vectors $b$ and $c$ will always be random vectors with independent Bernoulli $\pm 1$ entries.

Of particular interest is the case $N = n K$ for some $K \in \N$ and
$\Omega = \{ K, 2K, \hdots, nK\}$. Then the application of $S_\Omega^b$ and
$T_\Omega^c$ corresponds to (periodic or non-periodic) convolution with the sequence $b$ (or $c$, respectively)
followed by a downsampling by a factor of $K$. This setting was studied numerically 
in \cite{babadutrwa06} by Tropp et al.\ (using orthogonal matching pursuit instead of $\ell_1$-minimization). 
Also of interest is the case $\Omega = \{1,2,\hdots,n\}$ which was investigated in \cite{bahanorawr07,bahanora08} by Bajwa et al., who showed
that the restricted isometry constant of $T_\Omega^c$ satisfies 
$\delta_{\sparsity} \leq \delta$ with high probability (w.h.p.) provided $n \geq C_\delta \sparsity^2 \log(N/\sparsity)$. 
As a byproduct of the proof of our main result we give an alternative proof 
that 
$\delta_\sparsity \leq \delta$ holds w.h.p. under
the condition
$n \geq C \delta^{-2} \sparsity^2 \log^2(N)$. However, we strongly believe that
this bound is not optimal due to the quite pessimistic 
quadratic scaling in $\sparsity$. Our main result shows that one can achieve
recovery w.h.p. by $\ell_1$-minimization, if $n \geq C \sparsity \log^2(N)$.


In the following recovery theorem 
we use a random partial circulant or Toeplitz matrix 
$A^b_\Omega$ or $T^c_\Omega$ in the sense that the entries of the vector $b$ or $c$
are independent Bernoulli $\pm 1$ random variables. Furthermore, the signs of the non-zero
entries of the $\sparsity$-sparse vector $x$ are chosen at random according to a Bernoulli distribution as well. In contrast to previous work
\cite{bahanorawr07,babadutrwa06} 
$\Omega$ is allowed to be an arbitrary 
subset of $\{1,\hdots,N\}$ of cardinality $n$.

\begin{Theorem}\label{thm:main}
Let $\Omega \subset \{1,2,\hdots,N\}$ be an arbitrary (deterministic) 
set of cardinality $n$. 
Let $x\in\R^N$ be $\sparsity$-sparse such that the signs of its non-zero entries are Bernoulli $\pm 1$ random variables. Choose $b \in \R^N$ to be a random vector whose entries are $\pm 1$ Bernoulli variables. 
Let $y = S^b_\Omega x \in \R^n$. There exists a constant $C > 0$ such that
$$
n \geq C \sparsity \log^3(N/\epsilon)
$$  
implies that with probability at least $1-\epsilon$
the solution of the $\ell_1$-minimization problem \eqref{l1} coincides with $x$.

The same statement holds with $T^c_\Omega$ in place of $S^b_\Omega$ where $c \in \R^{2N-1}$
is a random vector with Bernoulli $\pm 1$ entries.
\end{Theorem} 
Ignoring the $\log$-factor the necessary number
of samples ensuring recovery by $\ell_1$-minimization scales linearly with the sparsity $\sparsity$. The power $3$ at the $\log$-term can very likely be improved to $1$, and moreover, it seems also possible to remove the randomness assumption on the non-zero coefficients of $x$. We postpone such improvements as well as an investigation of the restricted isometry constants 
to possible future contributions.
The remainder of the paper is concerned with the proof of Theorem \ref{thm:main}. 

\section{Proof of Theorem \ref{thm:main}}

An essential ingredient of the proof is the following recovery theorem
for $\ell_1$-minimization due to Fuchs \cite{fu04} and Tropp \cite{tr05-1}.
For a matrix $A$ we denote by $a_\rho$ its columns and by $A_\Lambda$
the submatrix consisting only of the columns index by $\Lambda$.

\begin{Theorem}\label{thm:Tropp} Suppose that $y = A x$ for some 
$x$ with $\supp x = \Lambda$. 
If
\begin{equation}\label{recovery:cond}
|\langle A_\Lambda^\dagger \atom_\rho,  \sgn(x_\Lambda) \rangle | < 1 \quad \mbox{ for all } \rho \notin \Lambda\,,
\end{equation}
then $x$ is the unique solution of the Basis Pursuit problem (\ref{l1}).
Here, $A_\Lambda^\dagger$ denotes the Moore-Penrose
pseudo-inverse of $A_\Lambda$.
\end{Theorem}

A crucial step in applying this theorem is to show that
the $\ell_2$-norm of $\A_\Lambda^\dagger \atom_\rho$ in \eqref{recovery:cond} is small. To this end one expands
\begin{equation}\label{PseudoInvEst}
\|\A_\Lambda^\dagger \atom_\rho\|_2 \,=\, \|(\A_\Lambda^* \A_\Lambda)^{-1} \A_\Lambda^* \atom_\rho\|_2 \,=\, \|(\A_\Lambda^* \A_\Lambda)^{-1}\|_{2\to 2} \|\A_\Lambda^* \atom_\rho\|_2,
\end{equation}
where $\|\cdot\|_{2\to2}$ denotes the operator norm on $\ell_2$.
The second term can be estimated in terms of the coherence of $\A$,
which is defined to be the largest absolute inner product 
of different columns of $\A$, $\mu = \max_{\rho \neq \lambda} |\langle \atom_\rho, \atom_\lambda \rangle|$. Indeed,
\[
\|\A_\Lambda^* \atom_\lambda\|_2 =
\left( \sum_{\lambda \in \Lambda} |\langle \atom_{\lambda}, \atom_\rho \rangle|^2 \right)^{1/2}
\leq \sqrt{|\Lambda|} \mu.
\]
The coherence of a random Toeplitz or circulant matrix 
can be bounded as follows.

\begin{proposition}\label{prop:coherence} 
Let $\mu$ be the coherence of the random partial circulant matrix
$\frac{1}{\sqrt{n}} S_\Omega^b \in \R^{n\times N}$ or Toeplitz matrix $\frac{1}{\sqrt{n}} T_\Omega^c \in \R^{n\times N}$ where
$b$ and $c$ are Rademacher series and $\Omega$ has cardinality $n$. 
Then with probability
at least $1-\epsilon$ the coherence satisfies
\[
\mu \leq 4 \frac{\log(2N^2/\epsilon)}{\sqrt{n}}. 
\]
\end{proposition}
The proof is contained in Section \ref{sec:proof:coherence}.
This proposition easily implies the following
(probably non-optimal) estimate of the restricted
isometry constants of $S^b_\Omega$ or $T^c_\Omega$ contained also
in \cite{bahanora08} with a different proof. 

\begin{corollary}\label{cor_RIP} 
Let $\frac{1}{\sqrt{n}} S_\Omega^b, \frac{1}{\sqrt{n}} T_\Omega^c \in \R^{n\times N}$ be 
the randomly generated normalized partial circulant and Toeplitz matrix generated from
Rademacher series and $\delta_\sparsity$ be their restricted isometry constant. 
Assume that
\[
n \geq 16 \delta^{-2} \sparsity^2 \log^2(2N^2/\epsilon).
\]
Then with probability at least $1-\epsilon$ 
it holds $\delta_\sparsity \leq \delta$.
\end{corollary}
\begin{Proof} Combine the bound $\delta_\sparsity \leq (\sparsity-1) \mu$ (which easily follows from
Gershgorin's disk theorem) with the estimate above on the coherence of $A = \frac{1}{\sqrt{n}} S^b_\Omega$ or $A = \frac{1}{\sqrt{n}} T^c_\Omega$.
\end{Proof}

As suggested by \eqref{PseudoInvEst} we also 
need an estimate of the operator norm of the inverse
of $A_\Lambda^* A_\Lambda$. To this end we bound
the smallest and largest eigenvalue of this matrix.

\begin{Theorem}\label{thm:condition} 
Let $\Omega,\Lambda \subset \{1,\hdots,N\}$ 
with $|\Omega| =n$ and $|\Lambda| = \sparsity$. Let $b \in \R^N$ and
$c \in \R^{2N-1}$ be Rademacher series. Denote either
$A = \frac{1}{\sqrt{n}} S^b_\Omega$ or $A = \frac{1}{\sqrt{n}} T^c_\Omega$. Assume
\begin{equation}\label{cond:conditioning}
n \geq \tilde{C} \delta^{-2} \sparsity \log^2(4\sparsity/\epsilon),
\end{equation}
where $\tilde{C} = 4\pi^2 \approx 39.48$. Then with probability
at least $1- \epsilon$ the minimal and maximal eigenvalues
$\lambda_{\min}$ and $\lambda_{\max}$ of $A_\Lambda^* A_\Lambda$
satisfy
\[
1-\delta \leq \lambda_{\min} \leq \lambda_{\max} \leq 1+\delta.
\]
\end{Theorem}
Note that the above theorem holds for a fixed subset 
$\Lambda$ and random coefficients $b$ or $c$. It does not 
imply that for given $b$ or $c$ the estimate holds uniformly
for all subsets $\Lambda$, which would be equivalent to
having an estimate for the restricted isometry constants
of $\frac{1}{\sqrt{n}} S^b_\Omega$ or 
$\frac{1}{\sqrt{n}} T^c_\Omega$. (Note that taking a union
bound over all subsets $\Lambda$ would yield an estimate essentially worse than Corollary \ref{cor_RIP}.)

Now we are ready to complete the proof of Theorem 
\ref{thm:main} on the basis of Proposition
\ref{prop:coherence} and Theorem \ref{thm:condition}.
We proceed similarly as in \cite[Theorem 14]{tr06-2}.
Hoeffding's inequality states that
\begin{equation}\label{Bernstein}
\P\big(|\sum_{j} \epsilon_j a_j|\geq u \|a\|_2\big) \leq 2 e^{- u^2/2}.
\end{equation}
By our assumption on the random
phases $\epsilon_\lambda = \sgn(x_\lambda)$, the scalar 
product on the left hand side of (\ref{recovery:cond}) 
is precisely of the above form
with $a = A_\Lambda^\dagger a_\rho = (A_\Lambda^* A_\Lambda)^{-1} A_\Lambda^* a_\rho$. 
Theorem \ref{thm:condition} implies that the 
smallest eigenvalue of $A_\Lambda^* A_\Lambda$ is bounded
from below by $1-\delta$ with probability at least
$1-\epsilon$ provided condition \eqref{cond:conditioning}
holds; hence, $\|(A_\Lambda^* A_\Lambda)^{-1}\|_{2\to 2} \leq \frac{1}{1-\delta}$. 
Plugging this into \eqref{PseudoInvEst} yields 
\begin{equation}\label{equation:CombineEstimates1}
\|\A_\Lambda^\dagger \atom_\rho\|_2 \leq \frac{1}{1-\delta} \sqrt{\sparsity} \mu.
\end{equation}
Following Theorem \ref{thm:Tropp} 
the probability that recovery fails can be estimated by
\begin{align}
&\P\big(|\langle \A_\Lambda^\dagger \atom_\rho, R_\Lambda \sgn(x) \rangle| \geq 1
\mbox{ for some } \rho \notin \Lambda\big)\notag\\[.2cm]
&\leq \P\big(\,|\langle \A_\Lambda^\dagger \atom_\rho, R_\Lambda \sgn(x) \rangle| \geq 1
\mbox{ for some } \rho \notin \Lambda\ \big|
 \mu \leq \frac{\alpha}{\sqrt{n}} \notag\\ 
& \phantom{\leq \P\big( }
\mbox{\&}\, \lambda_{\min} \geq 1-\delta\,\big) 
+ \P\big(\mu > \frac{\alpha}{\sqrt{n}}\big) + \P\big(\lambda_{\min} < 1-\delta\big) \notag\\[.2cm]
&\leq \sum_{\rho \notin \Lambda} \P\big(\,|\langle \A_\Lambda^\dagger \atom_\rho, R_\Lambda \sgn(x) \rangle| \geq 1\ \big| \ \mu \leq \frac{\alpha}{\sqrt{n}} \, \mbox{\&}\, \lambda_{\min} \geq 1-\delta\,\big) \notag\\[.2cm]&\qquad+
\P\big(\mu > \frac{\alpha}{\sqrt{n}}\big) + \P\big(\lambda_{\min} < 1-\delta\big).\notag
\end{align}
Under the assumption $\mu \leq \frac{\alpha}{\sqrt{n}}$ 
equation \eqref{equation:CombineEstimates1} implies that 
for $u= \frac {(1-\delta)\sqrt n}{\alpha \sqrt{\sparsity}}$ 
we have $u\| A_\Lambda^\dagger a_\rho\|_2\leq 1$, 
so (\ref{Bernstein}) gives
\begin{align}\label{equation:CombineEstimates2}
&\P\big(\,|\langle \A_\Lambda^\dagger \atom_\rho, R_\Lambda \sgn(x) \rangle| \geq 1\ \big|\ \mu \leq \frac{\alpha}{\sqrt{n}}\,  \mbox{\&} \, \lambda_{\min} \geq 1-\delta\, \big)\notag\\
&\leq \ 2 \exp\left(-\frac{(1-\delta)^2}{2\alpha^2} \frac{n}{\sparsity}\right).
\end{align}
Setting  
$\alpha = 4 \log(2N^2/\epsilon) 
$
Theorem \ref{prop:coherence} yields
\[
\P(\mu \geq \alpha/\sqrt{n}) \leq \epsilon.
\]
Now we choose 
$\delta = 1/2$. 
Under condition (\ref{cond:conditioning}), which reads
\begin{equation}\label{cond2}
n \geq 4\tilde{C} \sparsity \log^2(\sparsity/\epsilon),
\end{equation}
we have $\P(\lambda_{\min} \geq 1-\delta) \leq \epsilon$.
Hence, under the above conditions we obtain
\begin{align}
&\P\big(|\langle \A_\Lambda^\dagger \atom_\rho, R_\Lambda \sgn(x) \rangle| \geq 1
\mbox{ for some } \rho \notin \Lambda\big) \notag\\
&\leq 2 N \exp\left(- \frac{1}{8\log^2(2N^2/\epsilon)}\frac{n}{\sparsity}\right) + 2\epsilon.\label{P_estimate1}
\end{align}
The first term is less than $\epsilon$ provided
$n \geq 8 \sparsity \log^2(2N^2/\epsilon)\log(2 N/\epsilon)$, or
\begin{equation}\label{cond1}
n \geq C_1 \sparsity \log^3(N/\epsilon)
\end{equation}
for a suitable constant $C_1$. 
Conditions (\ref{cond2}) and (\ref{cond1}) are both 
satisfied if
\[
n \geq C \sparsity \log^3(N/\epsilon)
\]
for a suitable constant $C$, in which case the 
probability that recovery by $\ell_1$-minimization is less than $3 \epsilon$.
This completes the proof.

\section{Non-commutative Khintchine inequalities}

Both the proof of Proposition \ref{prop:coherence}
as well as the proof of Theorem \ref{thm:condition}
are based on versions of the Khintchine inequality. 
Let us first state the non-commutative 
Khintchine inequality
due to Lust-Piquard \cite{lu86-1} 
and Buchholz \cite{bu01}, see also \cite{tr06-2}.
To this end we introduce Schatten class norms on matrices.
Denoting by $\sigma(A)$ the vector of singular values of a matrix $A$, the $S_p$-norm is
defined as
\[
\|A\|_{S_p} := \|\sigma(A)\|_p,
\]
where $\|\cdot\|_p$ is the usual $\ell_p$-norm, $1\leq p\leq \infty$.

\begin{Theorem}\label{Gauss_Khintchine} 
Let $(A_k)$ be a finite sequence of 
matrices of the same 
dimension and let $(g_k)$ be a sequence of independent standard 
Gaussian random variables. Then for $m \in \N$,
\begin{align}
&\left[ \E \left\|\sum_k g_k A_k\right\|_{S_{2m}}^{2m} \right]^{1/2m} \notag\\
&\leq B_m \max \left\{ \left\|\left(\sum_k A_k A_k^*\right)^{1/2}\right\|_{S_{2m}}, \left\| \left(\sum_k A_k^* A_k \right)^{1/2} \right\|_{S_{2m}} \right\},\notag
\end{align}
with optimal constant 
\[
B_m = \left( \frac{(2m)!}{2^m m!}\right)^{\frac{1}{2m}} 
\]
\end{Theorem}
Using the contraction principle for Bernoulli random
variables, see \cite[eq.\ (4.8)]{leta91}, we obtain the non-commutative
Khintchine inequality for Bernoulli random variables \cite{lu86-1}.

\begin{corollary}\label{cor_Kh} Let $(A_k)$ be a finite sequence of matrices of the same 
dimension and let $(\epsilon_k)$ be a sequence of independent Bernoulli $\pm 1$
random variables. Then for $m \in \N$,
\begin{align}
&\left[ \E \left\|\sum_k \epsilon_k A_k\right\|_{S_{2m}}^{2m} \right]^{1/2m} \notag\\
&\leq C_m \max \left\{ \left\|\left(\sum_k A_k A_k^*\right)^{1/2}\right\|_{S_{2m}}, \left\| \left(\sum_k A_k^* A_k \right)^{1/2} \right\|_{S_{2m}} \right\},\label{Bernoulli_Khintch}
\end{align}
with constant 
\[
C_m = \sqrt{\frac{\pi}{2}} \left( \frac{(2m)!}{2^m m!}\right)^{\frac{1}{2m}}. 
\]
\end{corollary}
In the scalar case the factor $\sqrt{\pi/2}$ can be removed. However,
it is not clear yet whether this is true also in the non-commutative
situation.

The following theorem extends the non-commutative Khintchine inequality
to a second order chaos variable. Its proof uses decoupling and Corollary \ref{cor_Kh}.

\begin{Theorem}\label{Bernoulli_Khintchine} Let $A_{j,k} \in \C^{r \times t}$, $j,k=1,\hdots,N$, be matrices with $A_{j,j} = 0$, $j=1,\hdots,N$. 
Let $\epsilon_k$, $k=1, \hdots, N$ be independent Bernoulli
random variables. Then for $m \in \N$ it holds
\begin{align}
&\left[ \E \left\| \sum_{j,k=1}^N \epsilon_j \epsilon_k A_{j,k} \right\|_{S_{2m}}^{2m}\right]^{1/2m} \notag\\
&\leq D_m \max \left\{ \left\|\left(\sum_{j,k=1}^N A_{j,k} A_{j,k}^*\right)^{1/2}\right\|_{S_{2m}},\right.\notag\\
& \phantom{\leq D_m \max \{\{ }
\left. \left\| \left( \sum_{j,k=1}^N A_{j,k}^* A_{j,k}\right)^{1/2}\right\|_{S_{2m}}, \|F\|_{S_{2m}}\right\},   \notag
\end{align}
where $F$ is the block matrix $F = (A_{j,k})_{j,k=1}^N$ and the
constant
\[
D_m \,=\, 2^{1/2m} 2\pi C_m^2 = 2^{1/2m} 2\pi \left(\frac{(2m)!}{2^m m!}\right)^{1/m}.
\] 
\end{Theorem}

At present it is not clear whether the term $\|F\|_{S_{2m}}$ can be omitted above. At least, there is
no a priori inequality between any of the terms in the maximum.
The proof of the theorem is based on the following
decoupling lemma, see \cite[Proposition 1.9]{botz87} or \cite[Theorem 3.1.1]{gide99}.

\begin{lemma}\label{lem_decouple} Let $\xi_j$, $j=1,\hdots,N$, be a sequence of independent
random variables with $\E \xi_j = 0$ for all $j=1,\hdots,N$. Let 
$A_{j,k}$, $j,k=1,\hdots,N$, 
be a double sequence of elements in a Banach space with norm 
$\|\cdot\|$, where $A_{j,j} = 0$ for all $j=1,\hdots,N$.
Then for $1\leq p < \infty$
\[
\E \left\| \sum_{j,k=1}^N \xi_j \xi_k A_{j,k} \right\|^p  
\leq 4^p \E \left\| \sum_{j,k=1}^N \xi_j \xi'_k A_{j,k} \right\|^p,
\]
where $\xi'$ denotes an independent copy of the 
sequence $\xi = (\xi_j)$.
\end{lemma}
 
{\sc Proof of Theorem \ref{Bernoulli_Khintchine}}. We apply Lemma \ref{lem_decouple} followed
by the non-commutative Khintchine inequality \eqref{Bernoulli_Khintch},
\begin{align}
E& := \E \left\| \sum_{j,k=1}^N \epsilon_j \epsilon_k A_{j,k} \right\|_{S_{2m}}^{2m}\notag\\
&\leq 4^{2m} \E_{\epsilon} \E_{\epsilon'} \left\| \sum_{j,k=1}^N \epsilon_j \epsilon_k' A_{j,k} \right\|_{S_{2m}}^{2m}\notag\\
&\leq 4^{2m} C_m^{2m} \E_{\epsilon} \max\left\{ \left\|\left(\sum_{k=1}^N B_k(\epsilon)^* B_k(\epsilon)\right)^{1/2} \right\|_{S_{2m}}^{2m}, \right.\notag\\
&\phantom{\leq 4^{2m} C_m^{2m} } \left.
\left\| \left(\sum_{k=1}^N B_k(\epsilon) B_k(\epsilon)^*\right)^{1/2}
\right\|_{S_{2m}}^{2m} \right\},\label{MaxEst}
\end{align}
where $B_k(\epsilon) \,:=\, \sum_{j=1}^N \epsilon_j A_{j,k}$. We define
\begin{align}
\widehat{A}_{j,k} = \left(0|\hdots|0|A_{j,k}|0|\hdots|0\right) \in \C^{r \times tN}\notag
\end{align}
where the non-zero block $A_{j,k}$ is the $k$-th one, and similarly
\begin{align}
\widetilde{A}_{j,k} = \left(0|\hdots|0|A_{j,k}^*|0|\hdots|0\right)^* \in \C^{rN \times t}.\notag
\end{align}
Then clearly 
\begin{align}
\widehat{A}_{j,k} \widehat{A}^*_{j',k'} = \left\{\begin{array}{cc} 0 &  \mbox{ if } k \neq k',\\
A_{j,k} A_{j',k}^* & \mbox{ if } k = k'.
\end{array} \right.,\label{wideAA}\\
\widetilde{A}_{j,k}^* \widetilde{A}_{j',k'} = \left\{\begin{array}{cc} 0 &  \mbox{ if } k \neq k',\\
A_{j,k}^* A_{j',k} & \mbox{ if } k = k'.
\end{array} \right.\notag
\end{align}
The Schatten class norm satisfies $\| A \|_{S_{2m}} = \|(A A^*)^{1/2}\|_{S_{2m}}$.
This allows us to verify that
\begin{align}
&\left\|\sum_{j=1}^N \epsilon_j \sum_{k=1}^N \widehat{A}_{j,k} \right\|_{S_{2m}}
\,=\, \left\|\left(\sum_{j,j'} \epsilon_j \epsilon_{j'} \sum_{k,k'} \widehat{A}_{j,k} \widehat{A}_{j',k'}^*\right)^{1/2}\right\|_{S_{2m}}\notag\\
&=\, \left\|\left(\sum_{j,j'} \epsilon_j \epsilon_{j'} \sum_k A_{j,k} A_{j',k}^* \right)^{1/2} \right\|_{S_{2m}}\notag\\
&=\, \left\|\left(\sum_k B_k(\epsilon) B_k(\epsilon)^*\right)^{1/2}\right\|_{S_{2m}}.\notag
\end{align} 
Similarly, we also verify that
\[
\left\|\left(\sum_k B_k(\epsilon)^* B_k(\epsilon)\right)^{1/2}\right\|_{S_{2m}} = \left\|\sum_{j=1}^N \epsilon_j \sum_{k=1}^N \widetilde{A}_{j,k}\right\|_{S_{2m}}.
\] 
Plugging the above expressions into (\ref{MaxEst}) we 
can further estimate 
\begin{align}
E &\leq 4^{2m} C_{2m}^{2m} \left(\E \left\| \sum_{j=1}^N \epsilon_j \sum_{k=1}^N \widehat{A}_{j,k}\right\|_{S_{2m}}^{2m}\right.\notag\\
& \phantom{\leq 4^{2m} C_{2m}^{2m} ((}  
\left. + \E \left\| \sum_{j=1}^N \epsilon_j \sum_{k=1}^N \widetilde{A}_{j,k}\right\|_{S_{2m}}^{2m}\right).\notag
\end{align}
Using Khintchine's inequality \eqref{Bernoulli_Khintch} 
once more we obtain
\begin{align}
E_1\,&:= 
\E \left\| \sum_{j=1}^N \epsilon_j \sum_{k=1}^N \widehat{A}_{j,k}\right\|_{S_{2m}}^{2m}\notag\\
&\leq C_m^{2m} 
\max\left\{ \left\| \left( \sum_j \widetilde{B}_j \widetilde{B}_j^*\right)^{1/2} \right\|_{S_{2m}}^{2m},\right.\notag\\
&\phantom{\leq C_m^{2m} \max \{\{} \left.
\left\| \left( \sum_j \widetilde{B}_j^* \widetilde{B}_j\right)^{1/2} \right\|_{S_{2m}}^{2m} \right\}, \notag
\end{align}
where $\widetilde{B}_j \,=\, \sum_{k=1}^N \widehat{A}_{j,k}$. Using (\ref{wideAA})
we see that 
\[
\sum_j \widetilde{B}_j \widetilde{B}_j^* = \sum_{k,j} A_{j,k} A_{j,k}^*.
\]
Furthermore, with the block matrix 
\begin{align}
F = \left(\begin{array}{cc}\widetilde{B}_1\\\widetilde{B}_2\\ \vdots \\
\widetilde{B}_N\end{array}\right) 
\,=\, \left(\begin{array}{cccc}
A_{1,1} & A_{1,2} & \hdots & A_{1,N}\\
A_{2,1} & A_{2,2} & \hdots & A_{2,N}\\
\vdots & \vdots & \vdots & \vdots \\
A_{N,1} & A_{N,2} & \hdots & A_{N,N} \\ 
\end{array}\right) \notag
\end{align}
we have
\[
\|(\sum_k \widetilde{B}_k^* \widetilde{B}_k)^{1/2}\|_{S_{2m}}^{2m}
\,=\, \|(F^* F)^{1/2} \|_{S_{2m}}^{2m} \,=\, \|F\|_{S_{2m}}^{2m}.
\]
Hence,
\[
E_1 \leq C_m^{2m} \max\left\{ \|(\sum_{j,k=1}^N A_{j,k} A_{j,k}^*)^{1/2}\|_{S_{2m}}^{2m}, \|F\|_{S_{2m}}^{2m}\right\}.
\]
As $\widetilde{A}_{j,k}$ differs from $\widehat{A}_{j,k}$ only by interchanging
$A_{j,k}$ with $A_{j,k}^*$ we obtain similarly
\begin{align}
&E_2 := \E \left\| \sum_{j=1}^N \epsilon_j \sum_{k=1}^N \widetilde{A}_{j,k} \right\|_{S_{2m}}^{2m}\notag\\
&\leq \max\left\{\left\|\left(\sum_{j,k=1}^N A_{j,k}^* A_{j,k} \right)^{1/2}\right\|_{S_{2m}}^{2m},\|F\|_{S_{2m}}^{2m} \right\}. \notag
\end{align}
Finally,
we obtain
\begin{align}
& E \leq 4^{2m} C_m^{2m} (E_1 + E_2) \notag\\
& \leq 2 \cdot 4^{2m} C_m^{4m} \max\left\{\left\|\left(\sum_{j,k=1}^N A_{j,k}^* A_{j,k} \right)^{1/2}\right\|_{S_{2m}}^{2m},\right.\notag\\
& \phantom{\leq 2 \cdot 4^{2m} C_m^{4m} }
\left.
\left\|\left(\sum_{j,k=1}^N A_{j,k} A_{j,k}^* \right)^{1/2}\right\|_{S_{2m}}^{2m},\|F\|_{S_{2m}}^{2m}\right\}.\notag
\end{align}
This concludes the proof. \hfill\qed

Repeating the above proof for the scalar case (which removes the factor
$\pi/2$ in the constant) and applying
interpolation (see \eqref{Stirling1} and \eqref{interpol1} below) 
yields the following
(compare also \cite[Proposition 2.2]{mcta86}).

\begin{corollary}\label{Scalar_Khintchine} Let $a_{j,k} \in \C$, $j,k=1,\hdots,N$ be numbers 
with $a_{j,j} = 0$, $j=1,\hdots,N$. 
Let $\epsilon_k$, $k=1, \hdots, N$ be independent 
Bernoulli $\pm 1$
random variables. Then for $2 \leq p < \infty$ it holds
\[
\left[ \E \left| \sum_{j,k=1}^N \epsilon_j \epsilon_k a_{j,k} \right|^{p}\right]^{1/p} 
\leq d_p \left(\sum_{j,k=1}^N |a_{j,k}|^2\right)^{1/2},
\]
where the constant
\[
d_p \,=\, 4^{1/p} (4/e) p.
\] 
\end{corollary}

\section{Proof of the coherence estimate}
\label{sec:proof:coherence}

Now we are equipped to provide the proof of Proposition
\ref{prop:coherence}. An inner product of two columns $s_i, s_\ell$ 
of the normalized matrix $\frac{1}{\sqrt{n}} S^b_\Omega$ has 
the form
\[
\langle s_i, s_\ell \rangle \,=\, \frac{1}{n} \sum_{r\in \Omega} b_{i - r \mod N} b_{\ell - r \mod N}
\,=\, \frac{1}{n} \sum_{j,k=1}^N b_{j} b_k a_{j,k}^{i,\ell}, 
\] 
where $a_{j,k}^{i,\ell} = 1$ if $(j,k) = (i - r \mod N, \ell-r \mod N)$
for some $r \in \Omega$
and $a_{j,k}^{i,\ell} = 0$ otherwise. Similarly, the inner product of the columns $t_i$ of the
normalized matrix $\frac{1}{\sqrt{n}} T^c_\Omega$ can be written as
$
\langle t_i, t_\ell \rangle = n^{-1} \sum_{j,k=-N+1}^{N-1} c_{j} c_k \tilde{a}_{j,k}^{i,\ell}
$
with $\tilde{a}_{j,k}^{i,\ell} = 1$ if $(j,k) = (i-r,\ell-r) \in \{1,\hdots,N\}^2$ for some $r \in \Omega$
and $0$ otherwise. Observe that $\sum_{j,k} |a_{j,k}|^2 = \sum_{j,k} |\tilde{a}_{j,k}|^2 = |\Omega| =n$.
Now let $b\in \R^N$ and $c \in \R^{2N}$ be Rademacher series.
Then Corollary \ref{Scalar_Khintchine} yields
\begin{align}
&n\left(\E |\langle s_i, s_j\rangle|^p\right)^{1/p}=  \left(\E |\sum_{j,k} b_j b_k a_{j,k}^{i,\ell}|^p\right)^{1/p}\notag\\ 
&\leq 4^{1/p} (4/e) p  \left(\sum_{j,k} |a_{j,k}|^2 \right)^{1/2}= 4^{1/p} (4/e) p \sqrt{n}\notag
\end{align}
for $p\geq 2$, and the same estimate holds for $\E |\langle t_i, t_j\rangle|^p$.
In order to complete the proof we use the following 
simple and well-known probability
estimate, see e.g.\ \cite{leta91,tr06-2}.
\begin{lemma}\label{lem_moment1}  Suppose $Z$ is a 
positive random variable satisfying
$(\E Z^p)^{1/p} \leq \alpha \beta^{1/p} p^{1/\gamma}$ 
for all $p_0 \leq p < \infty$ and some $\alpha,\beta,\gamma > 0$. 
Then for arbitrary $\kappa > 0$,
\[
\P(Z \geq e^\kappa \alpha u) \leq \beta e^{-\kappa u^\gamma}
\]
for all $u \geq p_0$.
\end{lemma}
\begin{proof}
By Markov's inequality we obtain
\[
\P(Z \geq e^\kappa \alpha u)
\leq \frac{\E Z^p}{(e^\kappa \alpha u)^p}
\leq  \beta \left(\frac{\alpha p^{1/\gamma}}{e^\kappa \alpha u}\right)^p .
\]
Choosing $p=u^\gamma$ yields the statement.
\end{proof}
Lemma \ref{lem_moment1} with the optimal choice $\kappa = 1$ yields
\[
\P(n |\langle s_i,s_\ell \rangle| \geq  4 \sqrt{n} u)
\leq 4 e^{- u}
\]
for $u\geq 2$. Taking the union bound over all possible pairs of 
different columns $s_i, s_\ell$ we obtain
\begin{align}
\P(\mu \geq  4 n^{-1/2} u) \leq 2 N^2 e^{-u}.\notag
\end{align}
Set the right hand side to $\epsilon$. 
Then the resulting $u = \log(2N^2/\epsilon) \geq 2$ since 
we may assume without loss of generality that $N\geq 2$.
We obtain
\[
\P \left(\mu \geq 4 \frac{\log(2N^2/\epsilon)}{\sqrt{n}}\right) \leq \epsilon. 
\]
The same holds for the coherence of 
$\frac{1}{\sqrt{n}} T_\Omega^c$.

\section{Proof of Theorem \ref{thm:condition}}

We introduce the elementary shift operators on $\R^N$, 
$(S_j x)_\ell = x_{\ell - j \mod N}$, $j=1,\hdots,N$, and
\[
(T_j x)_\ell = \left\{\begin{array}{ll} x_{\ell-j} & \mbox{if } 1 \leq \ell-j \leq N, \\
0 & \mbox{otherwise}, 
\end{array} \right. 
\]
for $j = -N+1,\hdots,N-1$, $\ell=1,\hdots,N$. 
Further, denote by $R_\Omega: \R^N \to \R^\Omega$ the operator
that restricts a vector to the indices in $\Omega$. Then we can write
\begin{align}
S_\Omega^b = R_\Omega \sum_{j=1}^N \epsilon_j S_j \quad \mbox{and} \quad 
T_\Omega^c  = R_\Omega \sum_{j=-N+1}^{N-1} \epsilon_j T_j, \notag
\end{align}
where $(\epsilon_j)$ is a Rademacher sequence.  Denote by $A$ either $\frac{1}{\sqrt{n}} S_\Omega^b$ or
$\frac{1}{\sqrt{n}} T_\Omega^c$. We need to prove a bound
on the operator norm of $X_\Lambda := A_\Lambda^* A_\Lambda - I_\Lambda$
where $I_\Lambda$ denotes the identity on $\R^\Lambda$.
We introduce $R_\Lambda^*: \R^\Lambda \to \R^N$ to be the extension operator that fills up a vector in $\R^\Lambda$ with zeros outside $\Lambda$. Further, we denote by
$D_j$ either $S_j$ or $T_j$. Observe that 
\begin{align}
&A_\Lambda^* A_\Lambda \, =\, \frac{1}{n} \sum_{j}\epsilon_j R_\Lambda D_j^* R_\Omega^* \sum_{k} \epsilon_{k} R_\Omega D_{k} R_\Lambda^*\notag\\
&=\, \frac{1}{n} \sum_{\substack{j,k\\j\neq k}} \epsilon_j \epsilon_{k} R_\Lambda D_j^* P_\Omega D_{k} R_\Lambda^* + \frac{1}{n} R_\Lambda \left(\sum_{j} D_j^* P_\Omega D_j\right) R_\Lambda^*, \notag
\end{align}
where $P_\Omega = R_\Omega^* R_\Omega$ denotes the projection operator which
cancels all components of a vector outside $\Omega$. Here and in the following the sums range either
over $\{1,\hdots,N\}$ or over $\{-N+1,\hdots,N-1\}$ depending on whether we consider
circulant or Toeplitz matrices. It is straightforward to check that 
\begin{equation}\label{sum_Khintch}
\sum_{j} D_j^* P_\Omega D_j = n I_N,
\end{equation}
where $I_N$ is the identity on $\R^N$. Since $R_\Lambda R_\Lambda^* = I_\Lambda$ we obtain
\[
X_\Lambda = \frac{1}{n} \sum_{j\neq k} \epsilon_j \epsilon_i R_\Lambda D_j^* P_\Omega D_k R_\Lambda^* = \frac{1}{n} \sum_{j \neq k} \epsilon_j \epsilon_k A_{j,k}
\]
with $A_{j,k} = R_\Lambda D_j^* P_\Omega D_k R_\Lambda^*$. Our goal is to apply
Corollary \ref{Bernoulli_Khintchine}. To this end we first observe that by \eqref{sum_Khintch}
\begin{align}
\sum_j A_{j,k}^* A_{j,\ell} \, & = \, R_\Lambda D_k^* P_\Omega \left( \sum_j D_j P_\Lambda D_j^*\right) P_\Omega D_\ell R_\Lambda^*\notag\\
&=\, \sparsity R_\Lambda D_k^* P_\Omega  D_\ell R_\Lambda^*. \notag
\end{align}
Using \eqref{sum_Khintch} once more this yields
\[
\sum_{j,k} A_{j,k}^* A_{j,k} = \sparsity R_\Lambda \left(\sum_{k} D_k^* P_\Omega D_k\right) R_\Lambda^* = \sparsity n R_\Lambda R_\Lambda^* = \sparsity n I_\Lambda.
\]
Since the entries of all matrices $A_{j,k}$ are non-negative we get
\begin{align}
& \|(\sum_{j\neq k} A_{j,k}^* A_{j,k})^{1/2}\|_{S_{2m}}^{2m}  = \Tr \left( \sum_{j \neq k} A_{j,k}^* A_{j,k}\right)^m \notag\\
&\leq \Tr \left(\sum_{j,k}
A_{j,k}^* A_{j,k}\right)^m = \Tr \left( \sparsity n I_\Lambda\right)^m 
 = \sparsity^{m+1} n^m,  \notag
\end{align}
where $\Tr$ denotes the trace.
Furthermore, since $A_{j,k}^* = A_{k,j}$ we have 
$\sum_{j\neq k} A_{j,k}^* A_{j,k} = 
\sum_{j\neq k} A_{j,k} A_{j,k}^*$.
Let $F$ denote the block matrix $F = (\tilde{A}_{j,k})_{j,k}$ where $\tilde{A}_{j,k} = A_{j,k}$ if $j \neq k$ and $\tilde{A}_{j,j} = 0$. Using once again that
the entries of all matrices are non-negative we obtain
\begin{align}
& \|F\|_{S_{2m}}^{2m} = \Tr \left[(F^*F)^m\right] \notag\\
& = \Tr \left[\sum_{\substack{j_1,j_2,\hdots,j_m\\k_1,k_2,\hdots,k_m}} \tilde{A}_{j_1,k_1}^* \tilde{A}_{j_1,k_2} \tilde{A}_{j_2,k_2}^* \tilde{A}_{j_2,k_3} \cdots \tilde{A}^*_{j_m,k_m} \tilde{A}_{j_m,k_1}\right]\notag\\
& \leq \Tr \sum_{k_1,\hdots,k_m} \left[\sum_{j_1} A_{j_1,k_1}^* A_{j_1,k_2} 
\cdots \sum_{j_m} A_{j_m,k_m}^* A_{j_m,k_1}\right]\notag\\
& = \sparsity^m \Tr \sum_{k_1,\hdots,k_m} \left[R_\Lambda D_{k_1}^* P_\Omega D_{k_2} R_\Lambda^* R_\Lambda D_{k_2}^* P_\Omega D_{k_3} R_\Lambda^* \cdots
\right.
\notag\\
& \phantom{=  \sparsity^m \Tr \sum_{k_1,\hdots,k_m} [}\left.
\cdots R_\Lambda D_{k_m}^* P_\Omega D_{k_1} R_\Lambda^*\right],\notag
\end{align} 
where we applied also \eqref{sum_Khintch} once more. Using the 
cyclicity of the trace and applying \eqref{sum_Khintch} another time, together with
the fact that $T_k = T_{-k}^*$ and $S_k = S_{N-k}^*$, gives
\begin{align}
&\|F\|_{S_{2m}}^{2m} \leq \sparsity^m \Tr \left[\sum_{k_1} D_{k_1} P_\Lambda D_{k_1}^* P_\Omega \sum_{k_2} D_{k_2} P_\Lambda D_{k_2}^* P_\Omega \cdots
\right. \notag\\ 
& \phantom{\|F\|_{S_{2m}}^{2m} \leq } 
\left. \cdots \sum_{k_n} D_{k_n} P_\Lambda D_{k_n}^* P_\Omega\right] 
= \sparsity^{2m} \Tr[ P_\Omega]
= n \sparsity^{2m}.\notag
\end{align}
Since by assumption (\ref{cond:conditioning}) $\sparsity \leq n$ it follows that 
\begin{align}
&\|F\|_{S_{2m}}^{2m} \leq \|(\sum_{j\neq k} A_{j,k}^* A_{j,k})^{1/2}\|_{S_{2m}}^{2m}\notag\\
& = \|(\sum_{j\neq k} A_{j,k} A_{j,k}^*)^{1/2}\|_{S_{2m}}^{2m} \,\leq\, n^m \sparsity^{m+1}.\notag
\end{align}
Using $\|X_\Lambda\| = \|X_\Lambda\|_{S_\infty} \leq \|X_\Lambda\|_{S_p}$ and 
applying the Khintchine inequality in 
Theorem \ref{Bernoulli_Khintchine} 
we obtain for an integer $m$
\begin{align}
&\E\|X_\Lambda\|^{2m} = \E \|A_\Lambda^* A_\Lambda - I_\Lambda\|^{2m} \leq 
\E \|A_\Lambda^* A_\Lambda - I_\Lambda\|_{S_{2m}}^{2m}\notag\\
&= \frac{1}{n^{2m}} \E \|\sum_{j\neq k} \epsilon_j \epsilon_k A_{j,k}\|_{S_{2m}}^{2m}
\leq\, 2 (2\pi)^{2m} \left(\frac{(2m)!}{2^m m!}\right)^{2} \frac{\sparsity^{m+1}}{n^m}\notag.
\end{align}
Stirling's formula gives 
\begin{equation}\label{Stirling1}
\frac{(2m)!}{2^m m!} = \frac{\sqrt{2\pi 2m} (2m/e)^{2m} e^{\lambda_{2m}}}{2^m \sqrt{2\pi m}
(m/e)^m e^{\lambda_m}} 
\leq \sqrt{2}\, (2/e)^m m^m,
\end{equation}
where $\frac{1}{12m+1} \leq \lambda_m \leq \frac{1}{12m}$.
An application of H\"older's inequality yields for $\theta \in [0,1]$ and an arbitrary random variable $Z$.
\begin{align}
\E |Z|^{2m+2\theta} & = 
\E [|Z|^{(1-\theta)2m} |Z|^{\theta (2m+2)}]\notag\\
& \leq (\E |Z|^{2m})^{1-\theta} (\E |Z|^{2m+2})^\theta.\label{interpol1}
\end{align}
Combining our estimates above gives
\begin{align}
&\E \|X_\Lambda\|^{2m+2\theta}
\leq (\E\|X_\Lambda\|^{2m})^{1-\theta} (\E \|X_\Lambda\|^{2m+2})^{\theta}\notag\\
&\leq 4 (2\pi)^{2m+2\theta} (2/e)^{2m+2\theta} m^{2m(1-\theta)} (m+1)^{2\theta(m+1)} \frac{\sparsity^{m+\theta+1}}{n^{m+\theta}}\notag\\
&\leq 4 \left(\frac{4\pi}{e}\right)^{2m+2\theta} (m+\theta)^{2m+2\theta}
 \frac{\sparsity^{m+\theta+1}}{n^{m+\theta}},\notag
\end{align}
where we used the inequality between the geometric and arithmetic mean in the third step.
In other words, for $p\geq 2$,
\[
\left(\E \|X_\Lambda\|^{p}\right)^{1/p} \leq \frac{2\pi}{e} \sqrt{\frac{\sparsity}{n}} (4\sparsity)^{1/p} p.
\]
An application of Lemma \ref{lem_moment1} with the optimal value $\kappa = 1$ 
yields
\[
\P\left(\|X_\Lambda\| \geq 2\pi \sqrt{\frac{\sparsity}{n}} u\right) 
\leq 4\sparsity e^{-u}
\]
for all $u\geq 2$. Setting the right hand side equal $\epsilon$ shows that 
$\|X_\Lambda\| \leq \delta$ with probability at least $1-\epsilon$ provided
\[
n \geq (2\pi)^2 \delta^{-2} s \log^2(4\sparsity /\epsilon).
\]
This completes the proof of Theorem \ref{thm:condition}.

\section*{Acknowledgements}

This work was supported by the Hausdorff Center for Mathematics, University of Bonn and
the WWTF project SPORTS (MA 07-004).



\end{document}